\documentclass[review,number,sort,compress,showpacs,1p]{elsarticle}

\usepackage{graphicx}
\usepackage{dcolumn}
\usepackage{bm}
\usepackage{natbib}
\usepackage[bookmarks=false]{hyperref}
\usepackage{amsmath}
\usepackage{amssymb}

\newcommand{\PC}{\textsc{The Pencil-Code }}
\newcommand{\EQ}{\begin{equation}}
\newcommand{\EN}{\end{equation}}
\newcommand{\EQA}{\begin{eqnarray}}
\newcommand{\ENA}{\end{eqnarray}}

\newcommand{\Eqs}[2]{Eqs.~(\ref{#1}) and~(\ref{#2})}

\newcommand{\Fig}[1]{Figure~\ref{#1}}
\newcommand{\Tab}[1]{Table~\ref{#1}}

\def\Rey{\mbox{\rm Re}}
\def\St{\mbox{\rm St}}

\newcommand{\xx}{\mbox{\boldmath $x$} {}}

\newcommand{\uu}{\mbox{\boldmath $u$} {}}

\newcommand{\vv}{\mbox{\boldmath $v$} {}}

\newcommand{\FF}{\mbox{\boldmath $F$} {}}

\newcommand{\SSSS}{\mbox{\boldmath ${\sf S}$} {}}

\begin{document}
\title{Particle impaction efficiency and size distribution in a MSWI super heater
tube bundle}

\author[se]{Nils Erland L. Haugen}
\ead{nils.e.haugen@sintef.no}

\author[se]{Steinar Kragset\corref{cor1}}
\ead{steinar.kragset@sintef.no}

\author[se]{Mette Bugge}
\ead{mette.bugge@sintef.no}

\author[rw]{Ragnar Warnecke}
\ead{ragnar.warnecke@gks-sw.de}

\author[mw]{Martin Weghaus}
\ead{mene@weghaus.net}

\cortext[cor1]{Corresponding author}
\address[se]{Sintef Energy Research, N-7465 Trondheim, Norway}
\address[rw]{GKS, Schweinfurt, Germany}
\address[mw]{Weghaus, Waldb\"{u}ttelbrunn, Germany}

\begin{abstract}
  Particle impaction in the super heater geometry found in the
  municipal solid waste incinerator (MSWI) of GKS in Schweinfurt,
  Germany, has been investigated. By using direct numerical
  simulations for the fluid flow, inertial particles coupled to the
  fluid through the classical Stokes' drag law have been tracked.
  Focus has been on the effect of flow velocity,
  and it is shown that decreasing the flow velocity will drastically
  decrease the impaction efficiency for some particle radii. Finally
  particle size distribution measurements are presented and used to
  find quantitative mass fluxes both on the front and the back side of
  the tubes in the super heater tube bundle.
\end{abstract}


 \maketitle

\section{Introduction}
The deposition of particles on cylinders in an array is an important
phenomenon in systems ranging from heat exchangers to fibrous filter
screens. The research effort is motivated in the former class of
applications by the need to minimize the total particle deposition
\cite{li2008, huang1996, loehden1989, yilmaz2000}, whereas the
opposite is true for the filter systems \cite{suneja1974,
schweers1994, muhr1976, kasper2009}. A third motive is to reduce
erosion caused by the constant bombardment of solid particles onto the
cylinder surfaces in advanced coal-fired combustors and fluidised
beds \cite{tu1998,morsi2004,tian2007}.

Heat exchangers in coal combustion equipments or biomass fired boilers
typically consist of cylinders or tubes arranged in bundles around
which the flue gas is flowing. To a various extent this gas will
always contain fly ash particles resulting from impurities and
inorganic material in the fuel. Ash particles in a molten or highly
viscous state tend to stick to surfaces on impaction, forming deposits
that cause problems in terms of corrosion, efficiency loss and high costs for
maintenance. The understanding of the fluid-particle flow and the
deposition mechanisms is crucial for the design of such devices
\cite{scharler2007}.

The impaction of particles onto the cylinders strongly depends on the
velocity field of the fluid in the vicinity of the surface. In the
present study direct numerical simulations (DNS)
in the sense that the Navier--Stokes equations have been solved
without the use of any kind of modelling,
are used in order to accurately resolve the boundary layers around the
cylinders. The particle motion is described in the Lagrangian
formalism, and the coupling with the fluid is through the Stokes'
drag.  This enables us to study where the deposits will form and how
it depends on the Reynolds and Stokes numbers.
All the equations have been solved in two dimensions since basically
all major flow variations are in the plane normal to the tube axes
for low and intermediate Reynolds numbers. The flow is nevertheless
considered a three dimensional flow, so that the simulation results
can be interpreted as per unit length in the direction of the tube
axes.
The geometry of interest has been the super heater of the municipal
solid waste incinerator (MSWI) of Gemeinschaftskraftwerk Schweinfurt
GmbH (GKS) in Germany \cite{gks2009},
from which actual measurements of the flue gas particle size
distribution are presented towards the end of section
\ref{sec:results}.

In the super heater section of a boiler several impaction mechanisms
might have an effect. These mechanisms are typically; inertial
impaction, thermophoresis, turbulent eddy diffusion and Brownian
motions.  In the current work all other impaction mechanisms than
inertial impaction have been neglected.  The reason for doing this is
twofold. Firstly inertial impaction is the most important impaction
mechanism at least for large particles, and secondly inertial
impaction is also the most general mechanism; by performing a very
fundamental study on this mechanism separately, it will be possible to
clearly distinguish the importance of the different mechanisms at a
later stage.

A general investigation with all the impaction mechanisms included, but under
several assumptions and approximations, has been made at GKS with a 
commercial CFD program \cite{warnecke09}.

\section{Equations}
The simulations are carried out using \PC \cite{pc2009}, where the
governing fluid equations are
\EQ
\label{mom}
\rho \frac{D \uu}{D t}=-\nabla P+\nabla \cdot (2\mu \SSSS)
\EN
and
\EQ
\label{cont}
\frac{D \rho}{D t}=-\rho \nabla \cdot \uu,
\EN 
where
$t$ is time,
$P$ is pressure
$\uu$ is velocity,
$\mu=\rho \nu$ is the dynamic viscosity,
$\nu$ is the kinematic viscosity and
$\rho$ is density.
\EQ
\frac{D}{D t}=\frac{\partial}{\partial t}+\uu\cdot\nabla
\EN
is the advective derivative and the  rate of strain tensor is 
\EQ
\SSSS = \frac{1}{2}\left( \nabla \uu + (\nabla \uu)^T\right)
-\frac{1}{3}\nabla \cdot \uu.  
\EN
The isothermal equation of state,
\EQ
P=c_s^2\rho,
\EN
is used, where $c_s$ is the speed of sound.  The set of equations
\Eqs{mom}{cont} are solved at every grid point for every time step. As
is seen in the above equations no models 
or filters are used. As a consequence all spatial and temporal scales
must be resolved by the simulation. This requirement has led to the
use of a very high resolution grid of $1024\times4096$ grid points in
two dimensions in order to resolve a domain of 0.2~m~x~0.8~m. Since no
filters are used, and the discretization scheme is of high order, all
the small scale kinetic energy is, as in nature, dissipated purely by
the molecular viscosity.

While the fluid equations are solved at predefined grid points, the particles
are tracked individually.
The particle velocity is evolved in time by
\EQ
\frac{d\vv}{dt}=\frac{\FF_D}{m_p},
\EN
while the particle position behaves as
\EQ
\frac{d\xx}{dt}=\vv,
\EN
where $m_p$, $\vv$ and $\xx$ are the mass, velocity and position of the particle, 
respectively. 
The force $\FF_D$ is the drag force, 
\EQ
\label{drag1}
\FF_D=\frac{m_p}{\tau_p}\left(\uu-\vv\right).
\EN
No other forces are included in this work.
The particle response time is
\EQ
\label{taup}
\tau_p
=\frac{Sd^2C_c}{18\nu(1+f_c)},
\EN
where
$f_c=0.15 \Rey_p^{0.687}$ is negligible for small particles and
$S=\rho_p/\rho$. 
The particle Reynolds number is $\Rey_p=(d|\vv-\uu|)/\nu$,
the particle diameter is $d=2r$ when $r$ is the particle radius,
\EQ
C_c=1+\frac{2\lambda}{d}(1.257+0.4 e^{-(1.1d/2\lambda)})
\EN 
is the Stokes--Cunningham factor, and $\lambda$ is the mean free path for
a typical molecule in the gas.

Here $\rho_p$ is the density of the particle.
Assuming all particles to be small enough in order to neglect $f_c$ and
at the same time much larger than the mean free path of a molecule yields
\EQ
\tau_p=\frac{Sd^2}{18\nu}
\EN
which is a unique number for a particular particle in a given flow. For a more
detailed description of the simulations and \PC see 
Haugen \& Kragset (2010)~\cite{haugen2010}.

\section{Results}
\label{sec:results}
\begin{figure}[p]
\centering\includegraphics[width=0.4\textwidth]{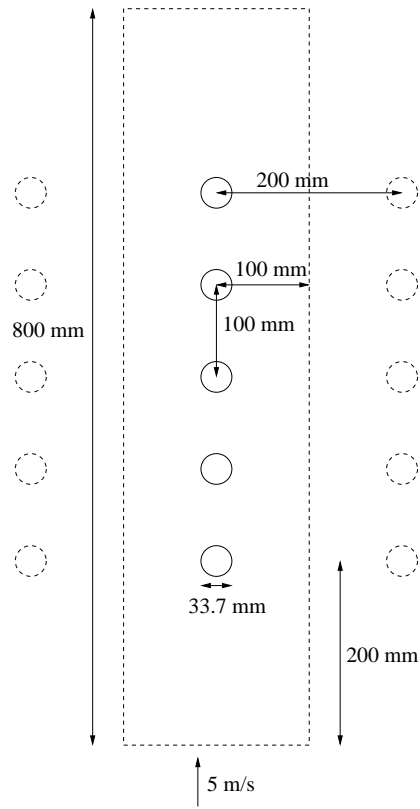}
\caption{Here the geometry of the first five tubes of the GKS plant is shown.
The dashed rectangle corresponds to the domain of the simulations, where
the lower boundary is the inlet and the upper boundary is the outlet. The side
boundaries are periodic, mimicking an infinite number of tube rows on each side
of the domain as illustrated by the two dashed tube rows on each side of the
rectangle. 
\label{geometry}}
\end{figure}

The super heater of the MSWI in the GKS plant \cite{gks2009} consists
of a non-staggered tube bundle where the center of the tubes are
separated by 200~mm in the transverse direction and by 100~mm in the
streamvise direction, see \Fig{geometry}.  Each tube has an outer
diameter of 33.7~mm and the mean velocity and temperature of the fluid
approaching the super heater is 5~m/s and 600$^\circ$~C,
respectively. This leads to a Reynolds number based on the mean
velocity and the cylinder diameter of $1685$ since a flue gas of
600$^\circ$~C yields a kinematic viscosity of approximately
$10^{-4}$~m$^2$/s. In all of the following we have set the
particle--fluid density ratio to $S=1000$.

The simulations presented here use periodic boundaries in the transverse
direction, this means that simulating a single tube row would represent an 
infinite number of tube rows. This is due to the fact that for periodic 
boundary conditions what goes out on one side is immediately inserted on the
other side of the domain. In the streamvise direction the five first 
tubes are simulated. Five tubes are chosen because initial calculations 
showed that the conditions after tube number five are essentially the same as
after a tube much further downstream. 

\subsection{Impaction efficiency as a function of Stokes number}
The main focus in the current work is the impaction efficiency $\eta =
N_{\rm{impact}}/N$ as a function of the Stokes and Reynolds
numbers. $N$ is here defined as the number of particles 
whose centers of mass
initially
are
moving in the direction of the tube bundle and $N_{\rm{impact}}$ is
the number of particles impacting on the tube. Since eddies can
deflect particles substantially, particles that initially were not
moving in the direction of the tube bundle may nevertheless impact,
causing $\eta$ to potentially exceed unity.
An impaction efficiency larger than one will in certain cases result
also if there is only a single target tube because of the particles'
finite extents~\cite{haugen2010}: Whenever a particle whose center of
mass moves closer to the tube than one particle radius, it is included
into $N_{\rm{impact}}$. In $N$ however, all particles are regarded as
point-like objects, and $N_{\rm{impact}} > N$ is consequently possible.

On impact the particle is removed from the simulation. 
Alternatively, a rebounding of the particles could have been allowed,
but to which degree this should happen would require knowledge of
materials specific parameter such as the sticking coefficient. A total
removal of the particles is therefore conveniently used, although
the opposite extremum could equally well have been chosen.
Furthermore
the Stokes number is given by
\EQ
\label{St1}
\St=\frac{\tau_p}{\tau_f}
\EN
and the Reynolds number
\EQ
\Rey=\frac{uD}{\nu},
\EN
where
\EQ
\tau_f=\frac{D}{2u}
\EN
is the fluid relaxation time and
$D$ is tube diameter.

\begin{figure}[p]
\centering\includegraphics[width=1.0\textwidth]{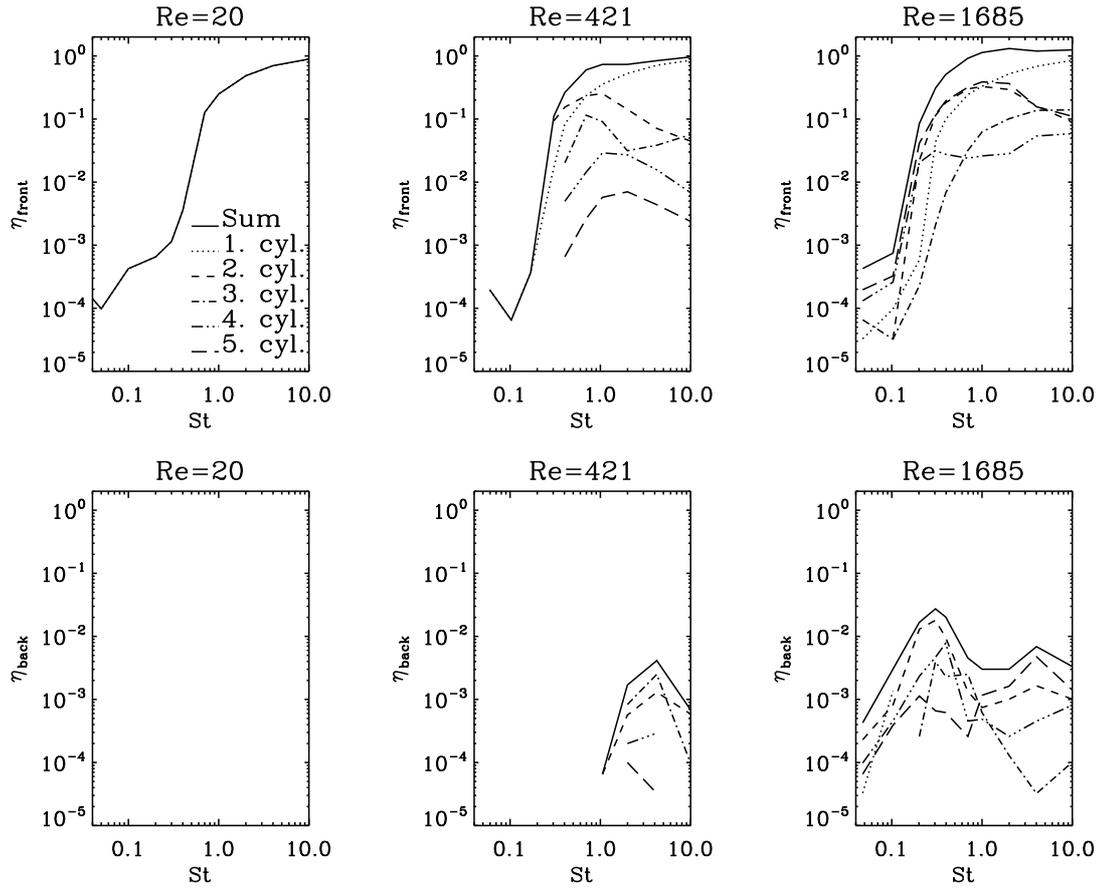}
\caption{Front side (upper plots) and back side (lower plots) 
impaction efficiency as a function of Stokes number
for different Reynolds numbers. Here both the impaction efficiencies for 
the sum of all the five tubes (solid line) and for each individual tube
are shown.
\label{fig1}}
\end{figure}

In the upper row of \Fig{fig1} the impaction efficiency on the front side of
the tubes is shown as a
function of Stokes number. For the low Reynolds number simulation it is 
seen that the only tube that experiences impaction is the first one. This is reasonable
as a Reynolds number of only 20 is sub-critical to the existence of 
von Karman eddies,
and consequently there are no fluid motion trying to force the particles 
towards the other cylinders as soon as they have passed the first one. Such a 
low Reynolds number is however not very relevant for a boiler as the heat
transfer rate from the fluid to the tubes downstream of the first tube will be
very low since the hot flow will pass between the tube rows without interacting
with the tubes. 

As the Reynolds number is increased to 421 (central upper plot in
\Fig{fig1}) the presence of von Karman eddies leads to particles
impacting also on the tubes downstream of the first tube.  Studying
the individual cylinders it can be seen that except for the initial cylinder,
which has an impaction efficiency which is monotonically increasing
with increasing Stokes number, all the downstream cylinders have peak
impaction efficiencies in the Stokes number range of 0.7 to 2. The
reason for this is that the particles with very large Stokes numbers
are not as efficiently affected by the von Karman eddies, and
consequently move straight on as they have passed the first tube. The
particles with very small Stokes numbers do not impact on the
downstream tubes simply because they follow the flow too well. The remaining
are the intermediately sized particles, having the
largest impaction efficiency as they are indeed affected by the von Karman
eddies, but they are still large enough in order to penetrate the
boundary layer around the tubes.

Increasing even further the Reynolds number to 1685 the impaction
efficiency of the downstream tubes is increased even more. The major
reason is the decrease in the boundary layer thickness as the
Reynolds number is increased. The boundary layer is effectively
working as a shield for the tubes against particle impaction. Another
reason of the increased impaction efficiency is the increased intensity
in the eddies generated by the cylinders. This is clearly seen in
\Tab{tab1} where the rms and maximum values of the transversal
velocities are shown. As expected the transversal velocities increase
with Reynolds number, and the maximum velocity is even larger than the
mean flow velocity for the largest Reynolds
numbers. This is reflected in the observation that for $\Rey=1685$
cylinders number 2, 4 and 5 all have larger impaction efficiency than 
the first cylinder for Stokes numbers smaller than 0.3.

\begin{table}[p]
  \caption{Root-mean-square and maximum transversal velocities for different
    Reynolds numbers.}
  \label{tab1}
  \begin{center}
    \begin{tabular}{r r r}
      \hline

 
      \multicolumn{1}{c}{$\Rey$} & 
      \multicolumn{1}{c}{$u_{\rm trans, rms}$} & 
      \multicolumn{1}{c}{$u_{\rm trans, max}$} \\ 

      \multicolumn{1}{c}{-} & 
      \multicolumn{1}{c}{m/s} & 
      \multicolumn{1}{c}{m/s} \\ 
     
      \hline
      20 &  0.3  &  2.9  \\
      421 &  1.2  &  8.1  \\
      1685 &  1.6  & 11.3  \\
      \hline
    \end{tabular}
  \end{center}
\end{table}

Focusing now on the back side impaction it is seen in the lower left
plot of \Fig{fig1} that there is no back side impaction for $\Rey=20$,
this is again reasonable because there are no von Karman eddies
generated for this Reynolds number and hence that there is no effect
which can force the particles to move towards the back side of the
cylinders.

For $\Rey=421$ (central lower plot of \Fig{fig1}) it is seen that
back side impaction occur for all cylinders except for the first one. 
The impaction efficiency is however relatively small, and it is largest around
$\St=4$. In particular it is worth mentioning that the relatively large
back side impaction for cylinder 3 around $\St=4$ is due to a very prominent
particle stagnation between cylinder 3 and 4. Particles tend to stay in 
this stagnation point for long times, but as they leave they gain velocity
opposite to the mean flow velocity in the direction of cylinder 3. As particles
with $\St=4$ are large particles they do not have any problems penetrating
the boundary layer on the back side of cylinder 3.

As the Reynolds number is increased to $\Rey=1685$ (lower right 
plot of \Fig{fig1}) the overall back side impaction show a bimodal behavior.
For the first cylinder we recover the same results as seen on a single cylinder
in Haugen \& Kragset (2010)~\cite{haugen2010}; 
that there is back side impaction only for Stokes
numbers smaller than $\sim 0.1$. The peak in impaction efficiencies at the 
smallest Stokes numbers is due to the second cylinder, while the large 
Stokes number peak is due to the last cylinder. It should be noted that the 
cause of the two peaks are fundamentally different. The peak at small 
Stokes numbers is caused
by particles being captured in the eddies behind a given cylinder and are 
then given a velocity in the direction of the back side of the cylinder by
this eddy. This is the same mode of back side impaction as is also found
for a single cylinder \cite{haugen2010}, and it leads to impaction
essentially on all angles on the back side. The second peak is caused
by particles being diverted by eddies formed by cylinders 
further up-stream in a neighbouring tube row. 
These particles will then approach the cylinder not from 
the front side but rather from the side, which then allows for particle 
impaction at an angle slightly larger than 90$^\circ$. It is indeed found that
for Stokes numbers larger than unity almost all the impaction on the back 
side occurs at angles between 90$^\circ$ and 110$^\circ$.

\begin{figure}[p]
\centering\includegraphics[width=1.0\textwidth]{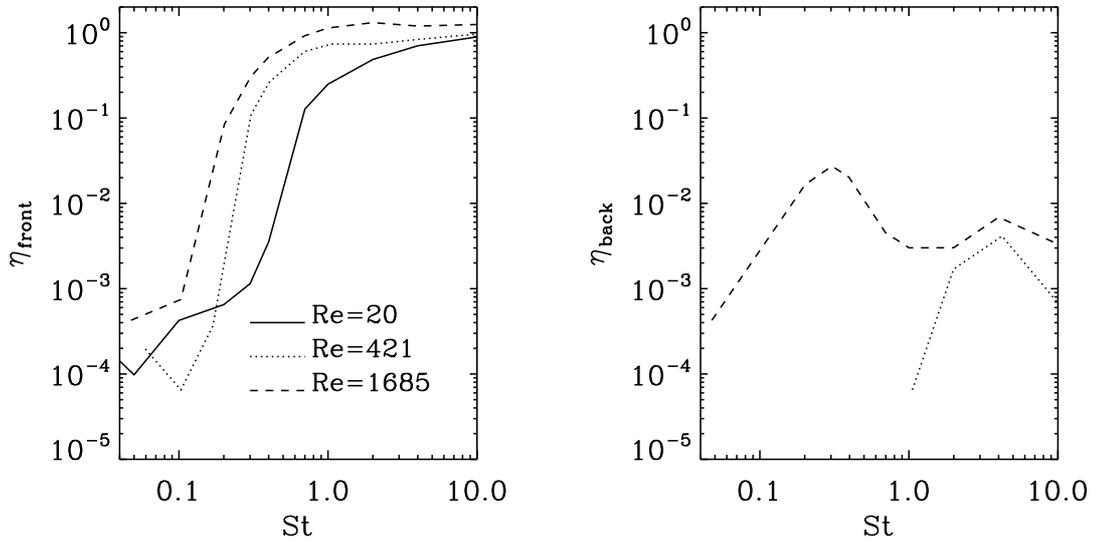}
\caption{Front side (left plot) and back side (right plot) impaction efficiency 
for different Reynolds numbers as a function of the Stokes number.
\label{fig2}}
\end{figure}

In \Fig{fig2} it is shown that the larger the Reynolds number the larger is
both the total front and total back side impaction efficiency. 
Here total refers to the sum of the impaction efficiencies for all cylinders
(which is indeed the same as the solid lines in \Fig{fig1}).
In  Haugen \& Kragset (2010)~\cite{haugen2010} 
it was shown that this was also the case for the front side impaction on 
a single cylinder for
Stokes numbers larger than $\sim 0.2$. For smaller Stokes numbers, however,
Haugen \& Kragset (2010)~\cite{haugen2010} 
found that the impaction efficiency was largest for
the small Reynolds numbers. Here the same trend is seen between Reynolds 
numbers of 20 and 421, but with a Reynolds number of 1685 the impaction
efficiency is always larger than for the smaller Reynolds numbers.
Regarding the back side impaction 
Haugen \& Kragset (2010)~\cite{haugen2010} found that
for a single cylinder there were no back side impaction for Stokes numbers
larger than 0.13. This is clearly not the case here.

\begin{figure}[p]
\centering\includegraphics[width=1.0\textwidth]{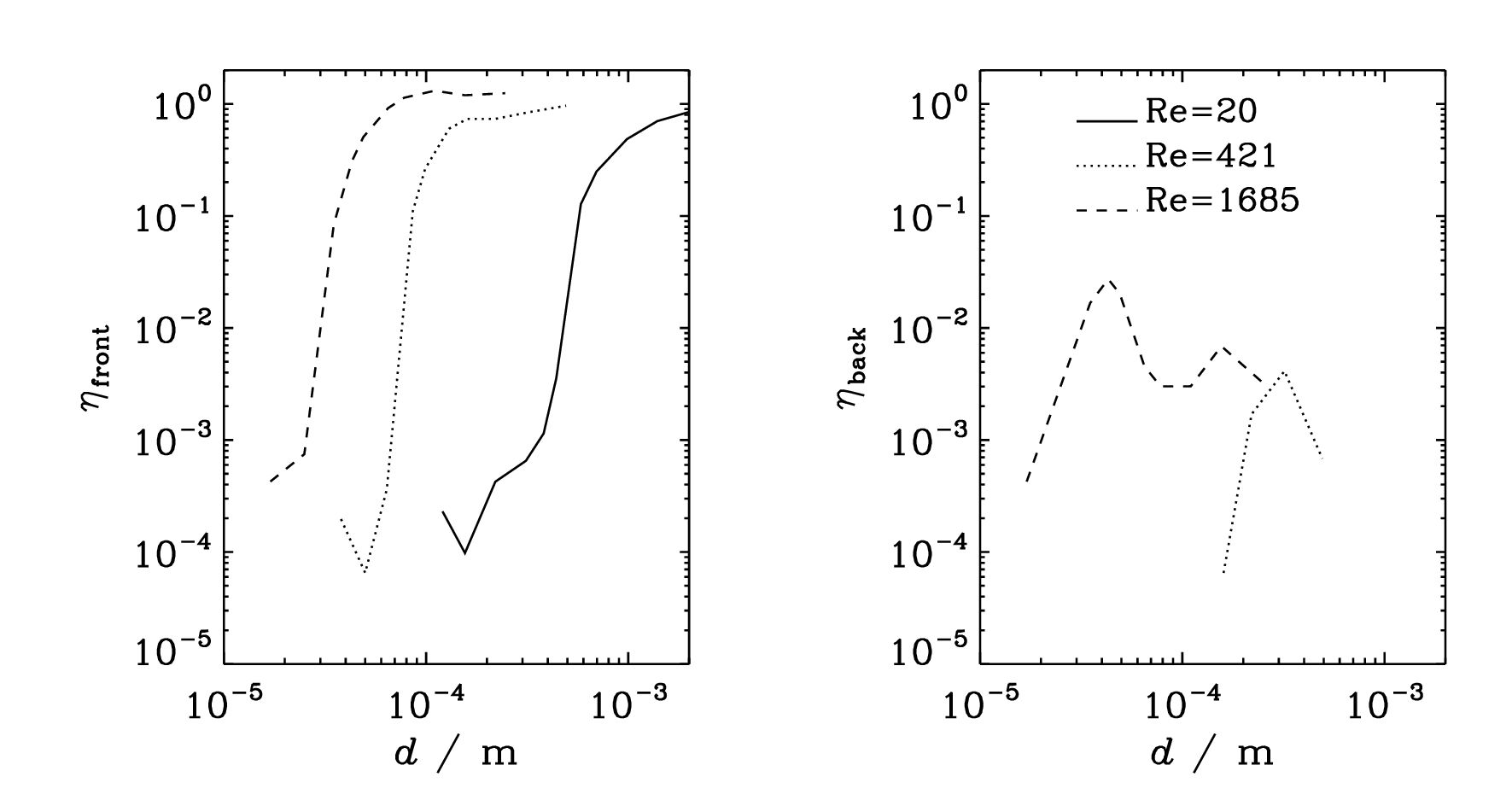}
\caption{Front side (left plot) and back side (right plot) impaction efficiency 
for different Reynolds numbers as a function of the particle diameter.
\label{fig3}}
\end{figure}

\subsection{Impaction efficiency as a function of particle diameter}
From the practical point of view, in an industrial boiler, it might be more
interesting to predict the impaction efficiency as a function of the particle
size than of the Stokes number. The latter is dependent on the super heater
geometry and dimensions, whereas more or less the same particle size
distribution will be emitted from the furnace no matter what is done with the
super heater section.

The Stokes number can be expressed as a function of the Reynolds number
by
\EQ
\St=\frac{d^2Su}{9\nu D}=\frac{d^2 S}{9D^2}\Rey
\EN
which shows that, if the Reynolds number is decreased by either decreasing the
velocity or increasing the viscosity, the Stokes number is also decreased.
Consequently it is seen that the Stokes number varies
linearly with the Reynolds number, such that a given particle size corresponds
to a smaller Stokes number when the Reynolds number is small. From 
\Fig{fig1} it is seen that the total front side impaction efficiency
is monotonically decreasing with decreasing Stokes number.
This means that for a given particle size the front side impaction 
efficiency will always decrease with the Reynolds number. This is shown more
clearly in \Fig{fig3} where
the total front and back side impaction efficiency is shown as a function of
particle size for the three different Reynolds numbers. It is seen that 
when impaction efficiency is plotted as a function of particle size
the difference between small and
large Reynolds numbers is even more prominent than when plotted 
against Stokes number. As an example the capture
efficiency of a particle with diameter of 60 $\mu$m is around $10^{-4}$ for 
$\Rey=421$ while it is around $0.7$ for $\Rey=1685$, which is a 
difference by a factor of more
the three orders of magnitude.

In this work the Reynolds number is changed by changing the viscosity, this
is however equivalent to inversely changing the velocity.

\subsection{Particle size distribution}

\begin{table}[p]
  \caption{Particle size distribution just before the super heater of the MSWI in
Schweinfurt, Germany}
  \label{tab2}
  \begin{center}
    \begin{tabular}{r r r r r r r r r r}
      \hline 

      \multicolumn{1}{c}{$d$} & 
      \multicolumn{1}{c}{$d_{\rm min}$}&
      \multicolumn{1}{c}{$d_{\rm max}$}& &
      \multicolumn{1}{c}{$\hat{\rho}_{p,i}$} &
      \multicolumn{1}{c}{$\phi_i$} &
      \multicolumn{1}{c}{$\eta_{\rm front}$}& 
      \multicolumn{1}{c}{$\dot{m}_i$} \\
      
      \multicolumn{1}{c}{$\mu$m} & 
      \multicolumn{1}{c}{$\mu$m} &
      \multicolumn{1}{c}{$\mu$m} & &
      \multicolumn{1}{c}{g/m$^3$} &
      \multicolumn{1}{c}{g/(s m$^2$)}&
      \multicolumn{1}{c}{-} &
      \multicolumn{1}{c}{g/(s m$^2$)} \\

      \hline
      2.84	&2.29  	      &3.52	   & & 0.006   &   790&0.0008 &  0.66\\
      4.37	&3.52	      &5.63	   & & 0.003   &   263&0.0007 &  0.20\\
      7.23	&5.63 	      &8.98	   & & 0.002   &    92&0.0006 &  0.05\\
      12.0	&8.98 	      &16	   & & 0.004   &    65&0.0003 &  0.02\\
      40	&25	      &63	   & & 0.023   &    91&0.2247 &    21\\
      89	&63	      &125	   & & 0.120   &   315&1.1967 &   377\\
      177	&125	      &250	   & & 0.220   &   296&1.2082 &   358\\
      354	&250	      &500	   & & 0.270   &   179&1.0105 &   181\\
      707	&500	      &1000	   & & 0.078   &    26&1.0209 &    27\\
      1400	&1000	      &2000	   & & 0.046   &     7&1.0415 &     8\\
      \hline
    \end{tabular}
  \end{center}
\end{table}

\begin{figure}[p]
\centering\includegraphics[width=0.8\textwidth]{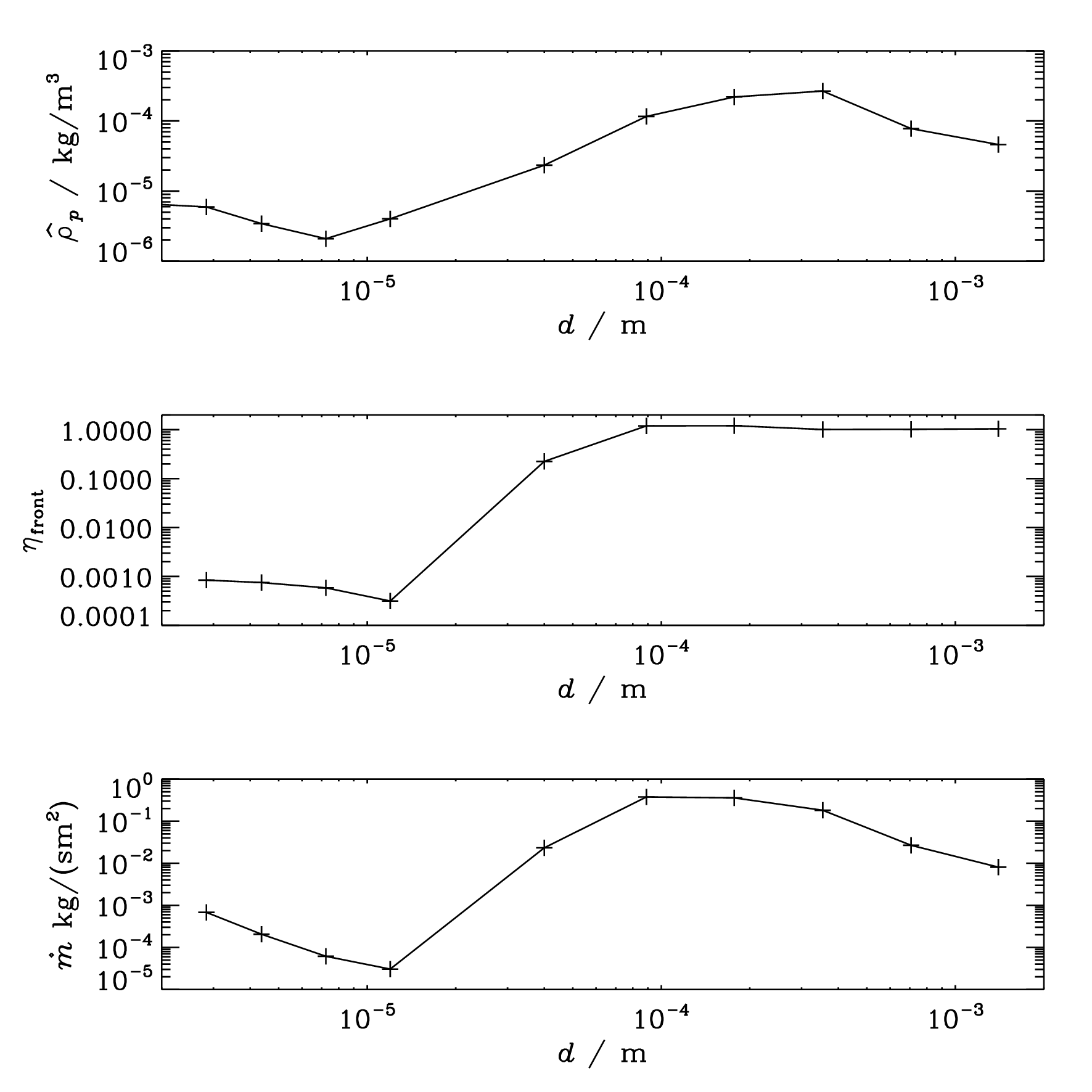}
\caption{The upper plot show the mass density for different particle sizes, while
the middle plot show the total impaction efficiency. Finally the lower plot show
the specific rate of mass impaction on the front side of the tubes, per tube
length, summed over all tubes in a tube row. 
The values result from a combination of
simulation data and measurements from the MSWI in Schweinfurt,
Germany.
\label{spec_mass_impact}}
\end{figure}

The particle impaction efficiency has been combined with measurements of
the flue gas particle size distribution in the super heater of the MSWI in
Schweinfurt, Germany, in order to provide quantitative predictions for
mass impaction on an actual tube bundle.  
In this subsection only the previously mentioned simulations 
with a Reynolds number of 1685 have been used.

In the measurements, the mass density $\hat{\rho}_p$ of particles in
the flue gas is size fractionated using different techniques for
coarse ($d \gtrsim 20 \mu$m) and fine ($d \lesssim 20 \mu\rm{m}$)
particles. For details, see \cite{deuerling2009}.
Note that this density is different from the internal particle density
$\rho_p$ and is defined as the mass of the particles (of a given size)
per fluid volume in which they are contained.
The results for different ranges of particle sizes have accordingly
been split into bins. That is, in particle bin $i$ the mean density
$\hat{\rho}_{p,i}$ of particles with diameters in the range between
$d_i$ and $d_{i+1}$ is given (see \Tab{tab2}).

As the time-step of the numerical simulations scales as $\sim d^2$ for
small particle diameters it is obvious that it is not practically
feasible to run simulations with extremely small particle
diameters. It was found that particle sizes less than a couple of
micrometers would require too much computer time. There are, however,
experimental results available for $\hat{\rho}_{p}$ down to
0.041~$\mu$m, but they have been omitted due to the restrictions
introduced by the simulations.

The average particle diameter\footnote{The average diameter is not an 
arithmetic mean
but is based on a fundamental knowledge of the measurements.} ${d}_i$ of
the interval $(d_{i,{\rm min}},d_{i,{\rm max}})$ will from now on be used as the 
representative diameter.

In the upper plot of \Fig{spec_mass_impact} the mean mass density is shown for 
particles in the range
$3\times 10^{-6}$~m to $1.4\times 10^{-3}$~m. 
Upstream of the first tube, the mean mass flux rate per tube length of 
particles in
bin $i$ flowing in the direction of the
tubes is
\begin{equation}
  \Phi_i = \hat{\rho}_{p,i} D u,
\end{equation}
and the corresponding specific mass flux rate for a particle diameter
within the bin is then 
\begin{equation}
  \phi_i \equiv \frac{\Phi_i}{\Delta d_{i}} =  \frac{\hat{\rho}_{p,i}
    D u}{\Delta d_i}, 
\end{equation}
where $\Delta d_i = d_{i+1}-d_i$. In the following all fluxes are measured 
per tube length.

The specific rate of mass impacting on the front side of the tubes in the 
tube bundle 
due to particles
with diameters belonging to bin $i$ is found as
\begin{equation}
  \dot{m}_i = \phi_i \eta_{\rm front}(d_i),
\end{equation}
and the results are shown in the lower plot of \Fig{spec_mass_impact}.
In the central plot of \Fig{spec_mass_impact} the front side impaction efficiency
is shown. This is essentially the same plot as the dashed line in \Fig{fig3}, 
but with
a slightly extended range towards smaller particles. 

\begin{figure}[p]
\centering\includegraphics[width=0.9\textwidth]{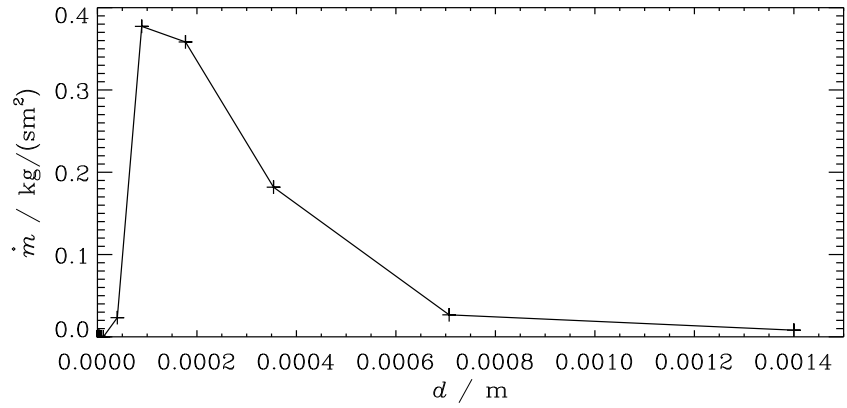}
\caption{
  The total specific rate of mass impaction on the front side of all the tubes
  in a tube row, per tube length.
\label{spec_mass_impact2}}
\end{figure}

Integrating $\dot{m}$ over particle diameter yields the total mass captured 
pr. time unit. 
In \Fig{spec_mass_impact2} $\dot{m}$ is consequently shown in a linear 
fashion such 
that the area under the graph corresponds to the total mass impaction rate.
From this it is seen that the major part of the mass deposition from 
particle impaction is due to particles in the range from $5\times 10^{-5}$~m 
to $7\times 10^{-3}$~m. So even though the impaction efficiency of
all particles with $d > 7\times 10^{-3}$~m is essentially unity, or even in
excess of unity, the total impacted mass due to these particles
is not very large because of their low abundance in the flow.

\begin{figure}[p]
\centering\includegraphics[width=0.9\textwidth]{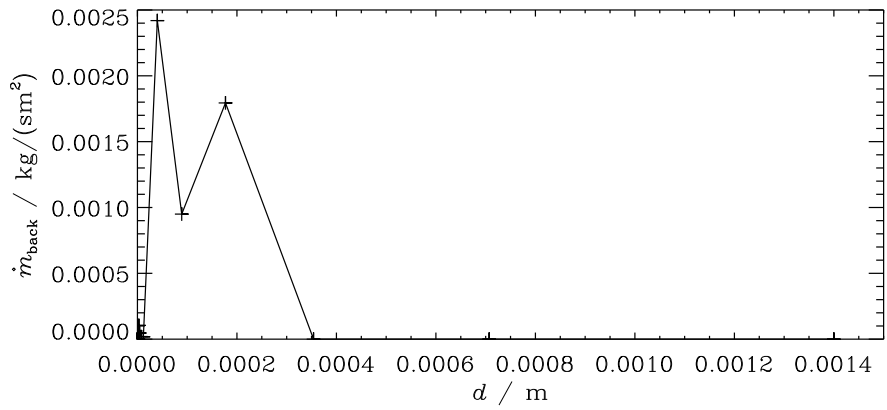}
\caption{The total specific rate of mass impaction on the back side of all 
  the tubes in a tube row, per tube length.
\label{spec_mass_impact2_back}}
\end{figure}

The specific rate of mass impacting on the {\it back} side of the tubes, 
$\dot{m}_{\rm back}$, is shown in \Fig{spec_mass_impact2_back}. It is seen that the
specific rate of back side mass impaction is first of all much smaller than 
what is
found on the front side. Furthermore the main peak of $\dot{m}_{\rm back}$  
is shifted slightly towards smaller particles compared to its front side 
counterpart.

\section{Conclusion}
In this work DNS has been used in order to accurately simulate particle
impaction on a tube bundle. The tube bundle has been set up to represent 
the super heater in the MSWI in Schweinfurt, Germany. It is found that
the particle impaction efficiency, both on the front and back side of the tubes,
is very dependent on Reynolds number. This is in particular true for the front
side capture of particles in the diameter range of 10-100~$\mu$m, where the
difference between $\Rey=1685$ and $\Rey=421$ is of several orders of 
magnitude. It must be highlighted here, though, that only the drag force
has been included. The inclusion of additional forces such as, Brownian 
motions and thermophoresis is expected to have a significant effect for small
particle sizes. Allowing for the flow entering the super heater tube bundle 
to be
turbulent may also have an effect on the results. 

It is shown that the back side impaction efficiency is significantly
increased for large Reynolds numbers. This is due to the increased
intensity in the eddies generated by the tubes as the Reynolds number
is increased.

Finally measurements of the particle size distribution found in the
MSWI in Schweinfurt are presented. These measurements are then used to
find quantitative results for the particle impaction rate for
different particle diameters. It is found that the largest part of the
mass impaction is found for particles in the range from $3\times
10^{-6}$~m to $1.4\times 10^{-3}$~m.

\section*{Acknowledgements}
This work has been part of the “NextGenBioWaste” project, 
co-funded by the European Commission under the Sixth Framework Programme.

\section*{Bibliography}
\bibliographystyle{elsarticle-num}

\end{document}